\documentclass[pra,twocolumn,showpacs,preprintnumbers,amsmath,amssymb,floatfix]{revtex4}
\usepackage{graphicx}
\usepackage{dcolumn}
\usepackage{bm}
\usepackage{graphicx}
\begin{document}
\title{Generation of multi-photon entanglement by propagation and detection }
\author{H. Hossein-Nejad}
\email{hnejad@physics.utoronto.ca}
\affiliation{Department of Physics, University of Toronto, 60 St. George St., Toronto, Canada}
\author{R. Stock}
\affiliation{Department of Physics, University of Toronto, 60 St. George St., Toronto, Canada}
\author{D. F. V. James}
\affiliation{Department of Physics, University of Toronto, 60 St. George St., Toronto, Canada}

\date{\today}

\begin{abstract}
We investigate the change of entanglement of photons due to
propagation. We find that post-selected entanglement in general
varies by propagation and, as a consequence, states with maximum bi-
and tri-partite entanglement can be generated from propagation of
unentangled photons. We generalize the results to n photons and show
that entangled states with permutation symmetry can be generated
from propagation of unentangled states. Generation of n-photon GHZ
states is discussed as an example of a class of states with the
desired symmetry.
\end{abstract}
\maketitle

\section{\label{sec:level1}Introduction}

It is well-known that the classical coherence properties of an
electromagnetic field vary due to propagation~\cite{MandelWolf}. The
most well-known  example of this is the increase of the spatial
coherence of the radiated field from an incoherent source upon
propagation~\cite{vanCittert:1934, Zernike:1938}; other examples are
the Wolf effect, the variation of the spectrum of light under
propagation~\cite{Wolf:1987, Wolf:1996}, and the change of
polarization of light under propagation~\cite{DFVJ:1994}.  It is
therefore reasonable to pose the question: can propagation alter the
\textit{quantum} correlations of a field?  In this paper, we study
the change of entanglement on propagation and answer this question
in the affirmative. A direct consequence of this result is
post-selective generation of polarization-entangled photons through
propagation of unentangled photons. For two photons this result has
been pointed out by Lim and Beige~\cite{Beige:2005} and is similar
in spirit to entanglement generation schemes in linear
optics~\cite{Weinfurter:2003,Fattal:2004, Pan:2001}, where erasure
of which-path information leads to creation of entanglement. The
scheme has also been implemented in reverse to generate entangled
atoms by detection of photons~\cite{Cabrillo:1999,Matsukevich:2008}.
Here, we extend these considerations to three photons and
demonstrate that propagation and post-selective measurement can be
used to create states with maximum genuine tri-partite
entanglement~\cite{Wootters:2000}. Furthermore, we show that a
generalization of this result leads to creation of $n$-photon
Greenberger-Horne-Zeilinger (GHZ) states.

Multi-photon entangled states have been generated for up to six
photon by down-conversion and linear-optics~\cite{Lu:2007} and are
of interest for optical quantum computing~\cite{Knill:2001,
Kok:2007}. Moreover, many-particle entangled states are a resource
in the one-way quantum computing paradigm~\cite{Raussendorf:2001,
Briegel:2009} which can be implemented advantageously in a linear
optics setting~\cite{Browne:2005, Prevedel:2007, Devitt:2008}.
Interferometric stability and beam-splitter alignment are major
obstacles in creation of larger entangled states by linear optics
techniques. It would therefore be desirable to create multi-photon
entanglement by simpler optical arrangements which may relax the
requirement for beam-splitter alignment and stability. The schemes
considered in this paper rely solely on free-space propagation and
detection; our study offers insight into generation and manipulation
of optical entanglement without the need for beam-splitters or
non-linear optical elements. One major drawback of the proposed
scheme is the exponential scaling with the number of qubits due to
the n-photon coincidence count. However, this deficiency can
possibly be overcome via classical interference, and is currently
under further investigation.

\section{Two photon case }
\subsection{Generation of Entanglement }

\begin{figure}[htbp]
\includegraphics[angle=270,width=75mm]{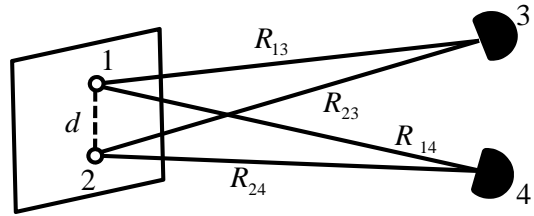}
\caption{Two photons are emitted from two pinholes 1 and 2 separated
by a distance $d$. Two detectors 3 and 4 are positioned in the
far-field of the photons. There are two possible interfering
processes by which the photons can reach the detectors. The
interference between these paths leads to generation of
entanglement.} \label{setnew}
\end{figure}

As an example of a simple situation in which propagation can change the quantum coherence of light, consider
the situation of Fig.\ref{setnew}. Suppose that one photon emerges from pinhole 1 in the polarization state
$a|H\rangle+b|V\rangle$, while a distance $d$ from pinhole 1 a second photon is radiated from pinhole 2 in
the state $c|H\rangle+d|V\rangle$. Their combined state is therefore
$(a|H\rangle + b|V\rangle) \otimes (c|H\rangle + d|V\rangle)$.
Two point-like detectors 3 and 4 are positioned in the far-field of
the pinholes. The photons reach the detectors after a time $r/c$,
where their joint state (polarization) is measured. We are only
interested in the events in which both detectors register a photon.
For this geometry, the state of the photons at the detectors is pure
and is given by
\begin{eqnarray}
& |\chi \rangle & = \frac{1}{\sqrt{N}} \{\frac{e^{ik(R_{13}+R_{24})}}{R_{13}R_{24}} (a|H\rangle+b|V\rangle)\otimes (c|H\rangle+d|V\rangle)  \nonumber \\&& + \frac{e^{ik(R_{23}+R_{14})}}{R_{23}R_{14}} (c|H\rangle+d|V\rangle)\otimes(a|H\rangle+b|V\rangle)\}
\label{eqn1}
\end{eqnarray}
where $k=\omega/c$ is the wavenumber of the photons, $R_{ij}$ is the distance between source $i$ and detector $j$ and $N$ is the normalization constant. There are two interfering processes in which both detectors could register a single count: photon 1 landing on detector 3 and photon 2 on detector 4; photon 1 arriving at detector 4 and photon 2 at detector 3. The two terms in the sum may be interpreted as the two possible paths taken by the photons. In the far field of the sources where $R_{13}R_{24}\simeq R_{23}R_{14}$, the state vector simplifies to
  \begin{eqnarray}
 |\chi \rangle & = & \frac{1}{\sqrt{N}} \{e^{-i \varphi} (a|H\rangle+b|V\rangle)\otimes(c|H\rangle+d|V\rangle) +  \nonumber \\&&  e^{i\varphi} (c|H\rangle+d|V\rangle)\otimes(a|H\rangle+b|V\rangle)\}
 \end{eqnarray}
where $\varphi = \frac{k}{2}(R_{14} + R_{23} - R_{13} - R_{24} )$
and we have assumed that the radial terms in the denominator of
Eq.~(\ref{eqn1}) are all of the same order in the far field and can
be absorbed in the normalization $N$. The entanglement of this state
can be quantified by concurrence~\cite{Wootters:1998} and is given
by
 \begin{equation}C = \frac{|(ad-cb)^2| }{2 \cos^2{\varphi} (|ac|^2+|bd|^2) +  (|ad|^2+|cb|^2+2|abcd|\cos{\varphi})} \label{C1}.\end{equation}
 As an example, if the photons begin in the state $|HV\rangle$, i.e. $a=d=1$ and $b=c=0$, one finds that the state at the detectors is $\frac{1}{\sqrt{2}}(|HV\rangle+|VH\rangle)$ and has entanglement of unity.

 This apparently counterintuitive result occurs because both photons are radiated into a large solid angle: whether a photon landing on a detector originated at 1 or 2 is therefore unknown prior to measuring its polarization. We then post-select \textit{only those events in which both detectors register a count}, projecting the detected state into a maximally entangled state.
\subsection{General Case }
\begin{figure}[htbp]
\includegraphics[angle=0, width=85mm]{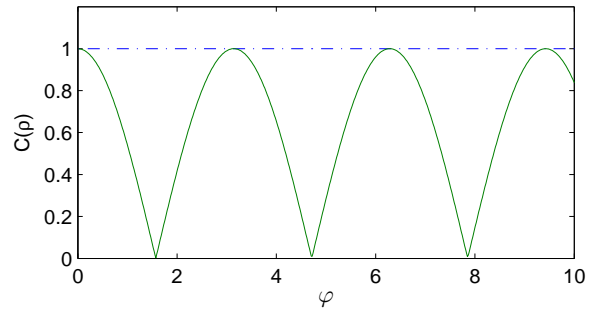}
\caption{ Far field entanglement versus the phase $\varphi$, for the
initial states $\frac{1}{\sqrt{2}}(|HV \rangle + i|VH \rangle)$
(solid green line) and $|HV\rangle$(dashed blue line). Concurrence
of the state generated from the initial state
$\frac{1}{\sqrt{2}}(|HV \rangle + i|VH \rangle)$ takes the simple
form $C(\rho)= |\cos(2\varphi)|$ in the far field.(Color online)}
\label{prop2}
\end{figure}

The above analysis can be generalized for an initial state of the
form  $\alpha|HH\rangle + \beta |HV\rangle+ \gamma |VH\rangle +
\delta |VV\rangle $,  i.e. an arbitrary pure state of the two
incident photons,  straightforwardly. The concurrence is now found
to be
\begin{equation}C = \frac{|\beta^2 +\gamma^2+ 2\beta\gamma\cos{2
\varphi}-4\alpha\delta \cos^2{\varphi}|}{1+\cos{(2\varphi)}(1-|\beta|^2-|\gamma|^2+2Re[\beta\gamma^{*}])}\label{general}
\end{equation}
As a first example, the entanglement of the initial state
$\frac{1}{\sqrt{2}}(|HV\rangle + |VH\rangle)$ is invariant under
propagation and the concurrence remains constant unity. The state
$\frac{1}{\sqrt{2}}(|HV\rangle + i|VH\rangle)$, however, generates a
state with concurrence of $|\cos{2\varphi}|$. Figure \ref{prop2}
shows a plot of concurrence versus the phase $\varphi$ for the
initial states $|HV\rangle$ and
$\frac{1}{\sqrt{2}}(|HV\rangle+i|VH\rangle)$. The entanglement of
the initial state $\frac{1}{\sqrt{2}}(|HV \rangle + i |VH \rangle)$
is destroyed and revived as the path difference between the two
possible paths for the photons to reach the detectors, is varied.

\subsection{Beam-Splitter Analogy }
\begin{figure}[htbp]
\includegraphics[width=47mm]{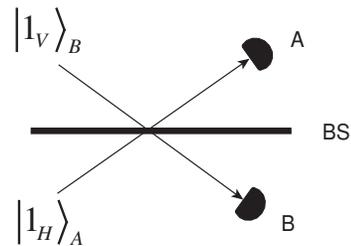}
\caption{The state emerging from the beam-splitter is entangled
provided that the photons are separated into two ports A and B.}
\label{bsp}
\end{figure}

Our scheme is reminiscent of standard entanglement generation schemes in linear optics
where two photons are incident on a beam-splitter
and the events in which
 the photons are separated into two different ports are post-selected (Fig.~\ref{bsp}).
 We use the notation $|n_i, m_j\rangle_{AB}\equiv |n_i\rangle_{A}|m_j\rangle_{B} $ to represent a
two-photon state in which $n$($m$) $i$-polarized ($j$-polarized)
photons are in the spatial mode $A$($B$), and each of the subscripts
$i$ and $j$ can take two possible values: $H$ or $V$. For the
incident state $|1_H, 1_V \rangle_{AB}$ the state emerging from the
beam-splitter takes the form ~\cite{Mandel:1988}
\begin{eqnarray} |\psi \rangle & = & \frac{1}{\sqrt{2}} \left( |1_H, 0 \rangle_{AB} + i |0, 1_H \rangle_{AB} \right) \otimes \nonumber \\&& \frac{1}{\sqrt{2}} \left( -i|1_V,0 \rangle_{AB} +  |0,1_V \rangle_{AB} \right)\end{eqnarray}
which is a product state and therefore not entangled. However, if we
select the events in which two photons separate into the
 two output ports A and B, the resultant state is the maximally
 entangled state  $ \frac{1}{\sqrt{2}} \left( |1_H,1_V \rangle_{AB} + |1_V,1_H\rangle_{AB} \right)$.
 Similarly, free space propagation acts like a beam-splitter, combining the state of
  the two photons. The lack of the Hong-Ou-Mandel effect ~\cite{Mandel:1988} in free space
    propagation, however, means that there is a subtle difference between the
    two schemes. To see this, consider a general state of the form
$\alpha|1_H,1_H\rangle_{AB} + \beta |1_H,1_V\rangle_{AB} + \gamma
|1_V,1_H\rangle_{AB} + \delta |1_V,1_V\rangle_{AB} $ incident on a
beam- splitter. Post-selecting the events at which each detectors
registers a count, we find the detected state to be $
\frac{1}{\sqrt{2}} \left( |1_H,1_V \rangle_{AB} +
|1_V,1_H\rangle_{AB} \right)$, provided that at least one of the
amplitudes $\beta$ or $\gamma$ is non-zero. If both $\beta$ and
$\gamma$ are zero, photon bunching prevents the photons from
arriving at different ports. This is in sharp contrast to the case
considered in the previous section where the entanglement of the
detected state has a strong dependency on the amplitudes of the
initial state.

Similar post-selective schemes have been utilized in linear
   optics quantum computing~\cite{Knill:2001,Kok:2007,Adami:1999} to generate
    entangled photons and perform logical operations or to induce
    effective non-linearities~\cite{Dowling:2003}. Since all such
    schemes rely on beam-splitters to erase the which-path
    information, their scalability is severely limited by
    interferometric stability; it would,
   therefore, be desirable to create multi-photon entangled states by
    simpler arrangements that do not rely on beam-splitters and/or
    Hong-Ou-Mandel effect.
\section{Two photon case: Emission by two atoms }
\begin{figure}[htbp]
\includegraphics[angle=270,width=70mm]{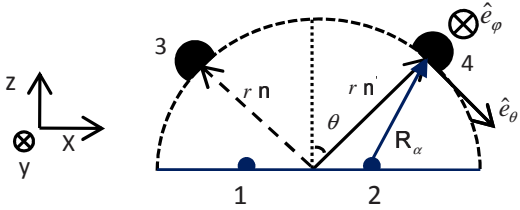}
\caption{Two single photon sources $1$ and $2$ and two detectors $3$
and $4$ are positioned in the xz-plane as shown. The detectors are
positioned symmetrically about the z-axis. The photon polarization
is measured along the unit vectors $\hat{e}_{\theta}$ and
$\hat{e}_{\phi}$.} \label{set}
\end{figure}

Having established the role of interference between paths in
entanglement creation/destruction, let us now consider a concrete
example where the photons originate from spontaneous emission of two
atoms. Apart from a means of implementing the scheme experimentally,
this calculation places the heuristic argument presented above on
firmer ground, and allows us to compute the entanglement at the
intermediate points from the near-field to the far-field of the
source. For a system consisting of two three-level atoms,
interacting with a quantized field, the Hamiltonian in the rotating
wave approximation is given by~\cite{M:Lehmber}
\begin{eqnarray}
& \hat{H}   = & \hbar \sum_{\bold{k}, \lambda} {\omega}_k {\hat{a}_{\bold{k}\lambda}}^{\dagger} {\hat{a}_{\bold{k}\lambda}} + \hbar {\omega}_{0} \sum_{i}^N \sum_{\lambda}^2 {\hat{s}_{\lambda i}}^{\dagger} {\hat{s}}_{\lambda i} +  \nonumber \\ && \hbar \sum_{\bold{k},i, \lambda} g_{i,\bold{k} , \lambda} \hat{a}_{\bold{k}\lambda} {{\hat{s}}_{\lambda i}}^{\dagger}+  c.c.
\end{eqnarray}
where $\sum_{i}$ indicates summation over the atoms,
$\sum_{\bold{k}}$ is the vector sum over the spatial field modes,
$\sum_{\lambda}$ is the summation over the two orthogonal
polarizations,   $\hat{s}_{\lambda i}$ (${\hat{s}_{\lambda
i}}^{\dagger}$) is the atomic lowering (raising) operator acting on
the $i^{th}$ atom  and corresponding to a transition with a
$\lambda$ polarized photon, ${\hat{a}_{\bold{k}\lambda}}$
(${\hat{a}^{\dagger}_{\bold{k}\lambda}}$) is the field annihilation
(creation) operator and $g_{i,\bold{k} , \lambda}$ is the coupling
constant. The two excited states are assumed to be degenerate with
${\omega}_0$ the atomic transition frequency. We choose to work in a
basis where the atomic operators correspond to linearly polarized
photons. Solving the Heisenberg equation of motion, $i \hbar
\dot{\hat{s}}_{\lambda i}(t)=[\hat{s}_{\lambda i}(t), \hat{H}]$, we
arrive at the following differential equation for the slowly varying
amplitude ${\hat{A}}_{\alpha i} = {\hat{s}}_{\alpha i} e^{i
\omega_{0}t}$,
\begin{eqnarray}
\frac{\partial {\hat{A}}_{\alpha i}}{\partial t} = i \frac{\gamma}{2} \sum_j M_{ij} {\hat{A}}_{\alpha j}(t)
\end{eqnarray}
where the matrix $M_{ij}$ describes the interaction between two atoms separated by $d$,
\begin{equation}
M_{ij} =
\left\{
\begin{array}{lr}
0&i=j.\\
\frac{c  }{d\omega_0} e^{i\omega_0 d/c} &i \neq j
\end{array}
\right.
\end{equation}
The solution to the differential equation is given by
\begin{eqnarray}
{\hat{s}}_{\alpha i}(t) = e^{-i \omega_0 t} \sum_{p,n} {\zeta_i}^{(p)} {\zeta_n}^{(p)} e^{i \gamma \alpha_p t/2} e^{-\gamma (1+\beta_p) t/2} {\hat{s}}_{\alpha i}(0)
\end{eqnarray}
where ${\zeta_n}^{(p)}$ is the $n^{th}$ eigenvector of the
interaction matrix $M$, with eigenvalue $\alpha_p+i\beta_p$. In
physical terms $\alpha_p$ and $\beta_p$ are the frequency shift and
the decay shift due to super-radiant effects which are small if the
atoms are more than one wavelength of radiation apart
\cite{Dicke:1954} and may also be neglected. As a consequence of the
orthogonality of the eigenvectors~\cite{M:note}, the Heisenberg
equation for the atomic operators is reduced to $ {\hat{s}}_{\alpha
i}(t) = {\hat{s}}_{\alpha i}(0) e^{-i {\omega}_0 t }  e^{-\gamma t/2
}$ for $\alpha=1$ or $2$, which is equivalent to the solution of a
classical oscillating dipole. The electric field at detector $j$ in
the semi-classical approximation is given by the formula
\cite{Lehmberg:1970}
\begin{eqnarray}
&& \bold{\hat{E}_j}^{(+)}(r\bold{n_j},t)=  \frac{p {{\omega}_0}^2 }{4 \pi {\epsilon}_0 c^2} \sum^{2}_{l=1} \Bigg(  \frac{\bold{n_j} \times [ \bold{n_j} \times {\bold{\hat{s}}}_{l}(t^{'}_{lj})] }{R_{lj}} \nonumber \\
&& +~3 ( \bold{n_j} [ \bold{n.\hat{s}}_l(t^{'}_{lj}) - \bold{\hat{s}}_l(t^{'}_{lj}) ] ) \left( \frac{1}{{R_{lj}}^3} - \frac{ik}{{R_{lj}}^2}\right)\Bigg) \label{field}
\end{eqnarray}
where $j=\{3,4\}$, $R_{lj}$ is the distance between atom $l$ and
detector $j$, ${t^{'}_{lj}}$ is the retarded time,
$t^{'}_{lj}=t-R_{lj}/c$,  $\bold{\hat{s}_l}= (\hat{s}_{1l},
\hat{s}_{2l}, 0)$, $p$ is the amplitude of the dipole moment of the
atom and $\bold{n_j}$ is the unit vector in the direction of the
observer. We have chosen the coordinates such that
$\hat{s}_{1l}$($\hat{s}_{2l}$) is the component of the dipole moment
operator along the x-axis (y-axis) (Fig.~\ref{set}). We use
$|x\rangle$, $|y\rangle$ and $|g\rangle$ to denote the eigenstates
of the atomic operators, we can therefore write
$\hat{s}_{11}\hat{s}_{12}|xx\rangle = |gg\rangle$,
$\hat{s}_{21}\hat{s}_{12}|yx\rangle = |gg\rangle$ and so on. The
radiated field may be simplified for an observer in the far-zone;
however, since we are interested in computing the propagation of
entanglement, all the terms in Eq.~(\ref{field}) will be kept. We
compute the state vector of the post-selected two-photon state
reaching the detectors for symmetric detection angles. The
components of the electric field at detector $j$ decomposed along
the azimuthal unit vectors $\bold{e_{\theta}} =
(\cos{\phi}\cos{\theta}, \sin{\phi}\cos{\theta}, -\sin{\theta})$ and
$\bold{e_{\phi}} = (-\sin{\phi}, \cos{\phi}, 0)$ are given by
\begin{eqnarray}
\label{Ephi}&& \hat{E}_{\phi j}^{(+)}(r\bold{n_j}, t) = \frac{p {{\omega}_0}^2 }{4 \pi {\epsilon}_0 c^2}  \sum^{2}_{l=1} (\hat{s}_{y l}(t) \cos{\phi}  - \nonumber \\&& \hat{s}_{x l}(t) \sin{\phi})  \left( \frac{k^2}{R_{lj}} + \frac{1}{{R_{lj}}^3} -\frac{ik}{{R_{lj}}^2}  \right) e^{i k R_{lj}} \end{eqnarray}
\begin{eqnarray}\label{Etheta} && \hat{E}_{\theta j}^{(+)}(r\bold{n_j}, t) = \frac{p {{\omega}_0}^2 }{4 \pi {\epsilon}_0 c^2} \sum^{2}_{l=1} (\hat{s}_{y l}(t) \cos{\theta}\sin{\phi}   + \nonumber \\&& \hat{s}_{x l}(t) \cos{\phi}\cos{\theta}) \left( \frac{k^2}{R_{lj}} + \frac{1}{{R_{lj}}^3} -\frac{ik}{{R_{lj}}^2}  \right) e^{i k R_{lj}}
\end{eqnarray}
where $\hat{s}_{x l}\equiv \hat{s}_{1 l}$ and $\hat{s}_{y l}\equiv
\hat{s}_{2 l}$ and the retarded dipole moment operator has been
written as $\hat{s}(t-R_{lj}/c) = \hat{s}(t)e^{ikR_{lj}}$.
We can simplify the calculations by remembering that all detected
two-photon states are pure. A general (unnormalized) state reaching
the detector is therefore of the form  $|\chi\rangle = C_{\theta
\phi}|HV\rangle + C_{\phi \theta}|VH\rangle + C_{\theta \theta}|HH
\rangle + C_{\phi \phi}|VV\rangle$ where $|H\rangle$ and $|V\rangle$
represent the two linear polarizations in the detector basis, along
the unit vectors $\bold{e}_{\theta}$ and $\bold{e}_{\phi}$
respectively. From the theory of photo-detection~\cite{Glauber:1963}
we know that $|C_{\theta \phi}|^2 = \langle \psi|
\hat{E}^{(-)}_{\theta3}(r\bold{n_3})
\hat{E}^{(-)}_{\phi4}(r\bold{n_4})
\hat{E}^{(+)}_{\theta4}(r\bold{n_4})
\hat{E}^{(+)}_{\phi3}(r\bold{n_3})  |\psi\rangle$ where
$|\psi\rangle$ is the initial state of the atoms and, since $t=r/c$,
we have omitted the time dependency of the fields. The amplitudes of
the detected state are therefore given by
\begin{equation}
\label{CCoeff} C_{\eta \kappa} = \langle gg| \hat{E}^{(+)}_{\eta3}(r\bold{n_3}) \hat{E}^{(+)}_{\kappa4}(r\bold{n_4})|\psi\rangle\end{equation}
where $\eta$ and $\kappa$ can be $\theta$ or $\phi$ and $|gg\rangle$
is the ground state of the atoms.

As an example for the \textit{initial atomic states} $|\psi_1\rangle
= |xy\rangle$, $|\psi_2\rangle = \frac{1}{\sqrt{2}}(|xy\rangle +
|yx\rangle)$ and $|\psi_3\rangle =
\frac{1}{\sqrt{2}}(|xy\rangle+i|yx\rangle)$, where $|x\rangle $ and
$|y\rangle$ represent the state of an atom in terms of the direction
of its dipole moment, the detected two-photon states are
respectively given by
\begin{equation*} |\chi_{1}\rangle = \frac{1}{\sqrt{2}}(|HV \rangle + e^{-2ikd \sin \theta} |VH \rangle) \end{equation*}
\begin{equation*} |\chi_{2} \rangle =\frac{1}{\sqrt{2}}(|HV \rangle - |VH \rangle) \end{equation*}
\begin{equation*} |\chi_{3} \rangle =-\frac{1}{2} \left( 1 + i e^{- 2ikd \sin \theta}\right) |HV \rangle + \frac{1}{2} \left( i +  e^{- 2ikd \sin \theta}\right) |VH \rangle.  \end{equation*}

\begin{figure}[htbp]
\includegraphics[angle=0, width=85mm]{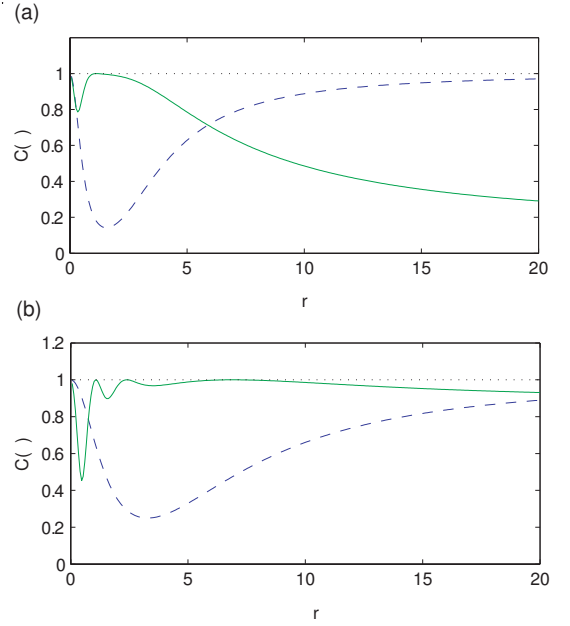}
\caption{Concurrence versus the detector position r,  in units of $\lambda/2 \pi$ for $\theta=45^{\circ}$. The initial atomic states are $|xy\rangle$(dashed blue line) and $\frac{1}{\sqrt{2}}(|xy\rangle+i|yx\rangle)$(solid green line). a) $kd=3.45$ (far-field minimum) and b) $kd=7$ (far-field maximum).(Color online)}
\label{prop}
\end{figure}
$|\chi_1\rangle$ and $|\chi_2\rangle$  have a concurrence of unity
and $|\chi_3\rangle$  has concurrence of $|\cos{(2\varphi)}|$, in
agreement with the heuristic treatment of the previous section. In
fact, if the initial state of the atoms is an arbitrary state of the
form $\alpha |xx\rangle + \beta|xy\rangle + \gamma|yx\rangle+
\delta|yy\rangle$, one arrives at Eq.~(\ref{general}) for
concurrence in the far-field.

We now compute the entanglement at all intermediate points as the
detectors are moved from the near-field to the far-field. For a
detection angle of $45^{\circ}$, the results are presented in
Fig.~(\ref{prop}). The behavior of the state  $|xy \rangle$ is
particularly intuitive to understand; at $r=0$ the detectors are at
the origin and concurrence is unity due to complete mixing of
photons. A ``near-zone'' minimum occurs at $r\sim\frac{d}{2}$ where
a detector is positioned close to each atom and the probability of a
photon captured by the farther detector is negligible. Concurrence
then recovers its far-field value as the detectors are moved further
apart. The state $|xy \rangle + i|yx \rangle$ also has unit
concurrence at $r=0$, but no subsequent ``near-zone'' minimum. This
is expected since the initial state was maximally entangled and the
entanglement is directly transferred from the atoms to the photons
if a detector is placed next to each atom. For $r<d/2$ a series of
interference fringes occur as the detectors are moved apart.
Concurrence recovers the far-zone value for $r>>d$.

\begin{figure}[htbp]
\includegraphics[angle=0, width=85mm]{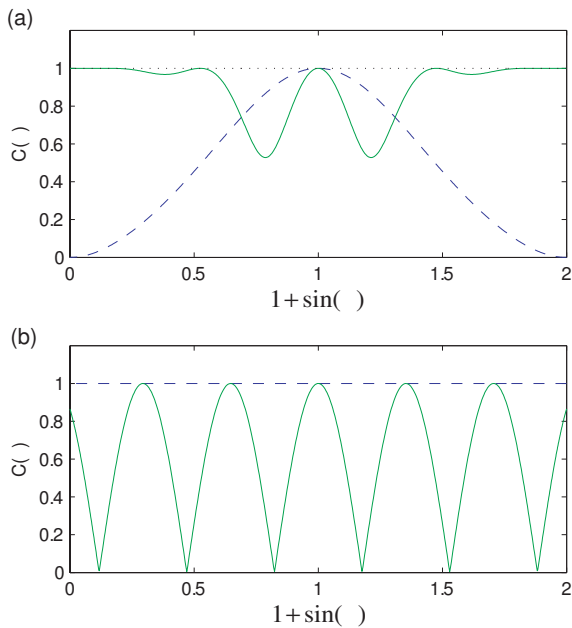}
\caption{Concurrence for symmetric detector positions for the
 initial atomic states $|xy\rangle$(dashed blue line) and
$\frac{1}{\sqrt{2}}(|xy\rangle+i|yx\rangle)$(solid green line). Both
detectors make an angle $\theta$ with the z-axis, where $-\pi/2 \leq
\theta \leq  \pi/2$. a) Concurrence in the near field for $kd=4.45$
and $r=2$. b) Concurrence in the far field for $kd=4.45$ and
$r=1000$.(Color online)} \label{polar}
\end{figure}

Finally to see the spacial variation of entanglement, we move the
detectors in the xz-plane, keeping them symmetric at all times, and
plot the variation of concurrence as a function of the detection
angle $\theta$ (Fig.~\ref{set}). Figure~\ref{polar}a. shows the
entanglement in the near-field. Figure~\ref{polar}b. is the
corresponding plot in the far-field where we have chosen the atomic
separation such that the state
$\frac{1}{\sqrt{2}}(|xy\rangle+i|yx\rangle)$ would have maximum
concurrence for a $\theta = 45^{\circ}$ detection angle. The state
$|xy \rangle$ generates maximally entangled photons for all
symmetric detector orientations. Similar fringe patterns are
observed at both extremes for the state $\frac{1}{\sqrt{2}}(|xy
\rangle + i |xy \rangle) $, confirming the previous observations
that altering the path difference between the photons can create
maxima and minima of entanglement. Our results are consistent with
the findings of Lim and Beige~\cite{Beige:2005} who study the
spacial variation of entanglement of formation for two dipole
sources, but do not discuss the variation of entanglement at the
intermediate points.
\section{n-photon GHZ-states}
\subsection{General Result}
The above observations give rise to the question: can propagation and
post-selection be used to create multi-photon entangled states? In
this section, we first state a general symmetry property of any
n-photon state that can be generated via propagation and
post-selection in our chosen geometry. We then show how
the initial conditions can be tailored to create n-photon GHZ states
up to local unitaries.
Finally we consider a three-photon GHZ state as a specific example.
\begin{figure}[htbp]
\includegraphics[width=75mm]{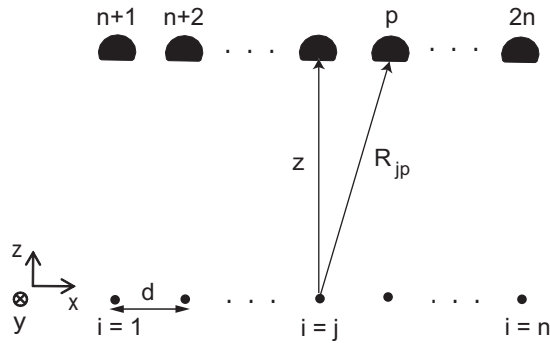} \caption{We consider an array of single photon
emitters prepared in the state $|\psi\rangle = \sum_{i_1...i_n}
D_{i_1... i_n}|i_1...i_n\rangle$. Polarization-resolving detectors
are positioned in the far-field of the atoms with the same spacing
as the atomic lattice.} \label{figure7}
\end{figure}
We assume an arrangement consisting of a one-dimensional array of
single photon emitters, separated by a distance $d$, with a single
photon detector directly above each emitter a distance $z$ away
(Fig.~\ref{figure7}). We consider an initial state of the form
$|\psi\rangle = \sum_{i_1...i_n} D_{i_1... i_n} |i_1...i_n\rangle$.
After propagation and post-selection, this produces a final state of
the form $|\chi\rangle = \sum_{i_1...i_n} C_{i_1... i_n}
|i_1...i_n\rangle$ where, each of the subscripts can take two
possible values, $H \equiv 0$ or $V \equiv 1$. We demonstrate that
states with permutation symmetry can be generated from propagation
of a suitable initial state, provided that the detectors are far
enough from the source. Local unitaries may need to be applied on
the photons before detection to create the desired entangled state.
A state $|\chi\rangle = \sum_{i_1...i_n} C_{i_1... i_n}
|i_1...i_n\rangle$ possess permutation symmetry if all amplitudes
$C_{i_1... i_n}$ with different permutations of the subscripts are
equal. The far-field condition is the key behind this result and
assumes a first order approximation for the distance between each
emitter and each detector; i.e. the phase difference between
different paths is neglected. This demands $\frac{n^2d^2}{2z^2}\ll1$
and becomes more difficult to satisfy for large $n$. The n-photon
GHZ state $|GHZ \rangle_n = \frac{1}{\sqrt{2}}
\left({|H\rangle}^{\otimes_n}  - {|V\rangle}^{\otimes_n} \right)$,
is an example of an state with the desired permutation symmetry; it
can thus be created by simple spatial propagation, if the amplitudes
of the initial state are chosen carefully. We demonstrate in
Appendix A that propagation of the \textit{separable} state
\begin{equation}
|\psi\rangle = \bigotimes_{l=1}^{n}\left(\sin\left(\frac{l\pi}{n}\right)|H\rangle + \cos\left(\frac{l\pi}{n}\right) |V\rangle\right) \label{magic}\end{equation}
generates the post-selected  \textit{entangled} state $|\chi\rangle$
up to global phases in the far-field such that
\begin{equation}
({\hat{{\mathcal{H}}} \hat{{\mathcal{S}}}^{\dagger})}^{{\otimes}^n}
|\chi\rangle = \frac{1}{\sqrt{2}} \left({|H\rangle}^{\otimes_n}  -
{|V\rangle}^{\otimes_n} \right)\label{unitaries}
\end{equation}
where $\hat{{\mathcal{H}}}$ is the Hadamard gate and
$\hat{{\mathcal{S}}}$ is the phase gate such that
$\hat{{\mathcal{S}}}|H\rangle = |H\rangle$,
$\hat{{\mathcal{S}}}|V\rangle = i|V\rangle$,
$\hat{{\mathcal{H}}}|H\rangle = \frac{1}{\sqrt{2}}(|H\rangle +
|V\rangle )$ and $\hat{{\mathcal{H}}}|V\rangle =
\frac{1}{\sqrt{2}}(|H\rangle - |V\rangle)$. Eq.~(\ref{magic})
corresponds to an initial state
 of the atoms such that the polarization for
 the n-th emitted photon simply forms an
 angle of $\frac{180^\circ}{n}$ with the horizontal.

Assuming emission into a $4\pi$ solid angle and a total detection
solid angle of $\Omega$, the probability of an $n$ photon
coincidence is $P_n = n!\eta^n$ for $\eta = \eta_q \eta_\Omega$,
where $\eta_q$ is the quantum efficiency of each detector and
$\eta_\Omega = \Omega/(4\pi n)$ (assuming that the area of each
detector scales as $\frac{1}{n}$). For large $n$, using Sterling's
formula for large factorials, one can show that the n-photon
coincidence scales as $P_n \sim \sqrt{2\pi n}\left(\eta_q \Omega
/(4\pi e)\right)^n$; in other words, it scales exponentially with
the size of the array. The main drawback of the scheme, is
therefore, the low probability of registering an n-photon
coincidence. One possibility of overcoming this obstacle is to use
classical interference to maximize the probability of detection at
the desired detector locations and will be investigated in a
subsequent publication.

\subsection{Three Photon GHZ state}

As a concrete example and to elucidate Eq.~(\ref{magic}) and
Eq.~(\ref{unitaries}), consider a general three-photon state of the
form $|\psi\rangle = \sum_{\eta \kappa \epsilon} D_{\eta \kappa
\epsilon} |\eta \kappa \epsilon \rangle$ which generates the state $
|\chi\rangle = \sum_{\eta \kappa \epsilon}C_{\eta \kappa
\epsilon}|\eta \kappa \epsilon\rangle$ in the far-field. By
generalizing Eq.~(\ref{CCoeff}) the amplitudes of the detected state
can be expressed as
\begin{equation}
\label{16} C_{\eta \kappa \epsilon} = \langle ggg| \hat{E}^{(+)}_{\eta4}(r\bold{n_4}) \hat{E}^{(+)}_{\kappa5}(r\bold{n_5}) \hat{E}^{(+)}_{\eta6}(r\bold{n_6})|\psi\rangle.
\end{equation}
This equation can be simplified for an observer in the far-field.
Inserting Eq.~(\ref{Ephi}) and Eq.~(\ref{Etheta}) into
Eq.~(\ref{16}) for $\phi=0$, we arrive at the far-limit of
Eq.~(\ref{16})
\begin{equation}C_{\eta \kappa \xi}= \langle ggg| \sum_{l m n}\sum_{\alpha \beta \gamma}D_{\alpha \beta \gamma}\hat{s}_{\eta l}\hat{s}_{\kappa m}\hat{s}_{\xi n}|\alpha \beta \gamma \rangle.\end{equation}
For the arrangement considered $R_{ji} \cong z$ for all $i$ and $j$
in the far-field and therefore all radial terms have dropped out.
The amplitude $C_{\eta \kappa \xi}$ is therefore the sum of six
terms: all cyclic and anti-cylic permutations of $\eta$, $\kappa$
and $\xi$:
\begin{equation}
\label{18} C_{\eta \kappa \xi} = D_{\eta \kappa \xi} + D_{\kappa \xi \eta } + D_{\xi \eta \kappa } + D_{\eta \xi \kappa } + D_{\kappa \eta \xi} + D_{\xi \kappa \eta  }
\end{equation}
This immediately proves, for example, that the state $|HHV\rangle$
generates a $W$ state in the far-field. To see how GHZ states are
created, one must demonstrate that propagation of the initial state
$|\psi\rangle = \bigotimes_{l=1}^{3}(\sin(\frac{l\pi}{3})
 |H\rangle + \cos(\frac{l\pi}{3}) |V\rangle)$ generates the state $|\chi\rangle
= \sum_{\eta \kappa \epsilon}C_{\eta \kappa \epsilon}|\eta \kappa
\epsilon\rangle = \frac{1}{2}(|HHV\rangle +|HVH \rangle +
|VHH\rangle - |VVV\rangle )$ in the far-field. To prove this it is
sufficient to show that a)~$C_{001}=C_{010}=C_{100}=-C_{111}$,
b)~$C_{110}=C_{101}=C_{011}=0$ . Both these criteria can readily be
verified from Eq.~(\ref{18}).

\begin{figure}[htbp]
\includegraphics[width=90mm]{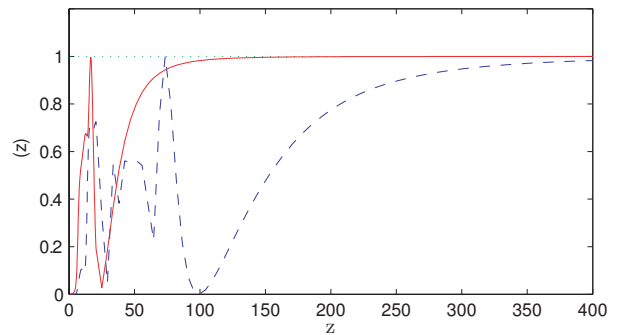}
\caption{Three-tangle versus the perpendicular detector position $z$
in units of $\lambda/2 \pi$ for the initial state described by
Eq.~(\ref{magic}) (n=3) for $d=\lambda$ (solid red line) and
$d=2\lambda$ (dashed blue line).} \label{three-tangle}
\end{figure}

Three qubit GHZ states have genuine tri-partite entanglement
\cite{Wootters:2000} and show maximum violation of three-qubit Bell
inequality~\cite{Ghose:2009}. Three-tangle is a measure of genuine
tri-partite entanglement and is defined to be~\cite{Wootters:2000}
\begin{equation}
\tau = {\cal{C}}_{1(23)}^2-{\cal{C}}_{12}^2-{\cal{C}}_{13}^2
\end{equation}
where ${\cal{C}}_{ij}$ is the concurrence between qubits $i$ and $j$
and ${\cal{C}}_{1(23)}$ measures the entanglement between qubit 1
and the joint state of qubits 2 and 3. Three-tangle is bounded
between 0 and 1 and states equivalent to GHZ states up to local
unitaries are characterized by a three-tangle of unity. Figure
\ref{three-tangle} is a plot of three-tangle versus distance for the
initial state $|\psi\rangle =
\bigotimes_{l=1}^{n}(\sin(\frac{l\pi}{n})
 |H\rangle + \cos(\frac{l\pi}{n}) |V\rangle)$ for two different initial atomic spacings. In the
far-field all phase information is washed out and three-tangle
approaches unity independent of the initial separation as expected.
\section{Conclusion}
In conclusion, we have demonstrated that entanglement of photons
emitted in a large solid angle can change on propagation. This
arises because in the far-zone all information about the origin of
the photons is lost and this leads to quantum
 mechanical interference between all possible paths to the
 detectors. We use concurrence and three-tangle as measures of two
 and three qubit entanglement and verify that both these quantities
 vary smoothly from near-field to far-field. We demonstrate that
 the propagation of the
 state $|\psi\rangle = \bigotimes_{l=1}^{n}(\sin(\frac{l\pi}{n})
 |H\rangle + \cos(\frac{l\pi}{n}) |V\rangle)$ generates n-photon
 GHZ states post-selectively. Our results appear to suggest that the chosen geometry is suitable for
generation of states with permutation symmetry; both n-photon W and
GHZ states fall into this category.

 The scheme
  may be realized experimentally using quantum dots or ions prepared in arbitrary
 initial states. The main drawback of the scheme is that
  the entangled photons are only accessible via post-selective
 or non-demolition measurements, which makes the scheme, in its current form, of limited use in
 practical generation of entanglement. However, generation of
 effective interactions between photons without the need for
 beam-splitters has obvious attractions in the design of linear
 optics quantum processors. One possibility of overcoming the
 post-selectivity criterion is using classical interference to
 maximize the probability of photon detection at the desired
 detector locations. This possibility is currently under investigation.
  \section*{Acknowledgments}
We thank Ignacio Cirac, Robert Prevedel, Aephraim Steinberg and
Andrew White for valuable discussions. This work was supported by
NSERC and the US Army Research Office.


  \section*{APPENDIX A: n-photon GHZ-states}
Here we present the proof that for a particular choice of the
initial state, the state reaching the detector is an n-photon GHZ
state up to local unitary rotations. We adopt the binary notation to
denote the polarization states,
\begin{equation}|0\rangle \equiv |H\rangle \end{equation}
\begin{equation}|1 \rangle \equiv |V\rangle.\end{equation}

\textit{Theorem}: For the specific geometry shown in
Fig.\ref{figure7} propagation of the state
\begin{equation}
|\psi\rangle = \bigotimes_{l=1}^{n}\left(\sin\left(\frac{l\pi}{n}\right)|0\rangle + \cos\left(\frac{l\pi}{n}\right) |1\rangle\right)
\end{equation}
post-selectively generates the state $|\chi\rangle$ up to global
phases in the far-field such that
\begin{equation}
({\hat{{\mathcal{H}}} \hat{{\mathcal{S}}}^{\dagger})}^{{\otimes}^n}
|\chi\rangle = \frac{1}{\sqrt{2}} \left({|0\rangle}^{\otimes_n}  -
{|1\rangle}^{\otimes_n} \right)\label{19}
\end{equation}
where the far-field is defined by the condition
$\frac{n^2d^2}{2z^2}\ll 1$, $\hat{{\mathcal{H}}}$ is the Hadamard
gate and $\hat{{\mathcal{S}}}$ is the phase gate such that
$\hat{{\mathcal{S}}}|0\rangle = |0\rangle$ and
$\hat{{\mathcal{S}}}|1\rangle = i|1\rangle $,
$\hat{{\mathcal{H}}}|0\rangle =\frac{1}{\sqrt{2}}(|0\rangle +
|1\rangle)$, $\hat{{\mathcal{H}}}|1\rangle =\frac{1}{\sqrt{2}}
(|0\rangle -|1\rangle) $.

\textit{Proof}: Let us assume initially that n is odd. The initial
state $|\psi\rangle$ and the final state $|\chi\rangle$ can be
rewritten as $|\psi\rangle = \sum_{i_1...i_n} D_{i_1... i_n}
|i_1...i_n\rangle$ and $|\chi\rangle = \sum_{i_1...i_n} C_{i_1...
i_n} |i_1...i_n\rangle$. By generalizing Eq.~(\ref{18}) we have
\begin{equation} C_{i_1... i_n} = \sum_{n!~perm.} D_{i_1... i_n} \end{equation}
where the summation is carried out over all $n!$ possible
permutations of the subscripts. The detected state possesses
permutation symmetry; meaning that all amplitudes with different
permutations of the same subscripts are equal. We use the notation
${\cal{C}}_{n,m}$ to denote the amplitude of an $n$-photon state in
which $m$ are in the state 1. We must show that the detected state
with no rotation is of the form
\begin{eqnarray}
({\hat{{\mathcal{S}}}}{\hat{{\mathcal{H}}}})^{{\otimes}^n}&\frac{1}{\sqrt{2}}&\left(|00...0\rangle-|11...1\rangle\right) =\nonumber\\
&& \frac{i}{\sqrt{N}} \sum_{i_1...i_n}B_{i_1...i_n}|i_1...i_n\rangle
\end{eqnarray}
where the amplitudes are symmetric with respect to permutations and
can be expressed as
\begin{equation}
{\cal{B}}_{n,m} = \left\{
\begin{array}{lr}
0 &m~even\\
(-1)^{\frac{m-1}{2}}&m~odd.
\end{array}
\right.
\end{equation}
The strategy is to prove that ${\cal{C}}_{n,m}\equiv
{\cal{B}}_{n,m}$. In order to do so we must show that:
a)${\cal{C}}_{n,m}=0$ for all even m, b)
$\frac{{\cal{C}}_{n,m}}{{\cal{C}}_{n,1}}= (-1)^{\frac{m-1}{2}}$ for
all odd $m$ with $n>m$.

\textit{Condition a}. The amplitudes $D_{i_1... i_n}$ can be written
as $ D_{i_1... i_n} = \Pi_{j=1}^{j=n} f(\frac{j\pi}{n}.i_j )$ where
the function $f(x.i_j )$ is defined to be
\begin{equation}
f(x.i_j ) = \left\{
\begin{array}{lr}
\sin(x)&\mbox{if}~~i_j=0\\
\cos(x)&\mbox{if}~~i_j=1
\end{array}
\right.
\end{equation}
We therefore have $C_{i_1... i_n} = \sum_{n!~perm.}\prod_{j=1}^{j=n}
f\left(\frac{j\pi}{n}.i_j\right)$. Since $i_n=1$ we can exclude the
last photon and write
\begin{equation}
C_{i_1... i_n} = \sum_{(n-1)!~perm.}\prod_{j=1}^{n-1}
f\left(\frac{j\pi}{n}.i_j \right)\label{26}\end{equation} which
vanishes if the product contains an odd number of cosines, or an
\text{even} number of 1s (including the $i_n$ photon). To see this
consider the amplitude ${\cal{C}}_{n,2}$. Expanding Eq.~(\ref{26})
we arrive at
\begin{eqnarray*}
&& {\cal{C}}_{n,2}  = \Bigg(
\sum_{i=1}^{\frac{n-1}{2}}\sin{(\frac{i\pi}{n})}\cos{(\frac{i\pi}{n})}\prod_{j=1,j\neq
i}^{\frac{n-1}{2}}\sin^2(\frac{j\pi}{n}) +  \\
&& \sum_{i=\frac{n+1}{2}}^{n-1} \sin{(\frac{i\pi}{n})}
\cos{(\frac{i\pi}{n})}\prod_{j=\frac{n+1}{2},j\neq
i}^{n-1}\sin^2{(\frac{j\pi}{n})}\Bigg)(n-2)!2!
\end{eqnarray*}
which vanishes since $\sin(\frac{i\pi}{n}) = \sin(\frac{n-i}{n}\pi)$ and
$\cos(\frac{i\pi}{n}) = -\cos(\frac{n-i}{n}\pi)$. This proves condition a.

\textit{Condition b.} We now write expressions for ${\cal{C}}_{n,1}$
and ${\cal{C}}_{n,m}$, excluding all vanishing terms in
Eq.~(\ref{26}) we obtain
\begin{eqnarray}
{\cal{C}}_{n,1}=(n-1)! \prod_{i=1}^{\frac{n-1}{2}}\sin^2{\left(\frac{i\pi}{n}\right)}
\end{eqnarray} and
\begin{eqnarray}
&&{\cal{C}}_{n,m} =  (-1)^{m'} (n-m)!m! \sum_{i_1>i_2>...i_{m'}}^{\frac{n-1}{2}}
\cos^2{\left(\frac{i_1\pi}{n}\right)} ...  \nonumber \\
&&\cos^2{(\frac{i_{m'}\pi}{n})} \prod_{k \neq i_{\alpha},
\alpha=1,..m'}^{\frac{n-1}{2}} \sin^2{\left(\frac{k \pi}{n}\right)}
\label{28}
\end{eqnarray}
where $m'=\frac{m-1}{2}$. The ratio of
${\cal{C}}_{n,1}$ and ${\cal{C}}_{n,m}$ is therefore given by
\begin{eqnarray}\frac{{\cal{C}}_{n,m}}{{\cal{C}}_{n,1}} =  (-1)^{\frac{m-1}{2}}   \sum_{i_1>i_2...>i_{m'}}^{\frac{n-1}{2}}\cot^2{\left(\frac{i_1\pi}{n}\right)}\cdots&& \nonumber \\
\cot^2{\left(\frac{i_{m'}\pi}{n}\right)}
\frac{(n-m)!m!}{(n-1)!}.\end{eqnarray} Condition b therefore
demands:
\begin{eqnarray}\label{nasty}
\sum_{i_1>i_2>...i_{m'}}^{\frac{n-1}{2}}\cot^2{(\frac{i_1\pi}{n})}\cdots\cot^2{(\frac{i_{m'}\pi}{n})} \equiv \frac{(n-1)!}{m!(n-m)!}
\end{eqnarray}
This equation can be proven by recalling that $t_r
=\cot^2{(\frac{r\pi}{2p+1})}$ for $r=1,2,...p $ are the distinct
roots of the $p$th degree polynomial~\cite{ProofsBOOK}
\begin{equation}Q(t):=\sum_{k=0}^{p}{2p+1 \choose 2k+1}(-1)^k t^{p-k}\label{polynom}\end{equation}
where ${n \choose k}=\frac{n!}{k!(n-k)!}$  . By applying Vi\'{e}te's
formula~\cite{Algebra} to Eq.~(\ref{polynom}) and making the
substitution $p=\frac{n-1}{2}$ and $m=\frac{m-1}{2}$ we arrive at
Eq.~(\ref{nasty}). This proves condition b.

If $n$ were assumed to be even from the outset, the upper limit of
the sum in Eq.~(\ref{nasty}) would have been $\frac{n}{2}-1$. The
proof for even n would have then demanded the substitution
$p=\frac{n}{2}-1$ in Eq.~\ref{polynom}.


\begin{thebibliography}{33}
\bibitem[{M:n()}]{M:note}
\bibinfo{note}{{Matrix $M$ is non-hermetian symmetric, the left and the right
  eigenvectors are therefore equal and we have $\sum_n \zeta_n^{(p)}
  \zeta_n^{(q)} = \delta_{p,q}$, $\sum_p \zeta_{n}^{(p)} \zeta_{m}^{(p)} =
  \delta_{n,m}$}}.
\end{thebibliography}

\end{document}